\documentclass[fleqn,usenatbib]{mnras}

\usepackage{newtxtext,newtxmath}
\usepackage[T1]{fontenc}

\DeclareRobustCommand{\VAN}[3]{#2}
\let\VANthebibliography\thebibliography
\def\thebibliography{\DeclareRobustCommand{\VAN}[3]{##3}\VANthebibliography}

\usepackage{graphicx}	% Including figure files
\usepackage{amsmath}	% Advanced maths commands
\usepackage{tikz}
\usepackage{natbib}
\usetikzlibrary{arrows.meta, positioning, calc}
\usepackage{pgfplots}
\usepackage{booktabs}
\pgfplotsset{compat=newest}
\usepackage{booktabs}
\usepackage{array}
\usepackage{multirow}
\title[Variational views for SSL in radio astronomy]{Variational views for self-supervised learning in radio astronomy}

\author[J.~Joseph Alphonse et al.]{
Johnny Joseph Alphonse,$^{1}$\thanks{E-mail: johnny.josephalphonse@postgrad.manchester.ac.uk (JJA)} and
Anna M. M. Scaife$^{1,2}$
\\
$^{1}$Jodrell Bank Centre for Astrophysics, University of Manchester, Oxford Road, Manchester M13 9PL, UK\\
$^{2}$Alan Turing Institute, Euston Road, London, UK\\
}

\date{Accepted XXX. Received YYY; in original form ZZZ}

\pubyear{\the\year{}}

\begin{document}
\label{firstpage}
\pagerange{\pageref{firstpage}--\pageref{lastpage}}
\maketitle

% Abstract of the paper
\begin{abstract}
Modern astronomical surveys are producing progressively larger and more complex datasets, making traditional supervised approaches that rely on extensive labelled catalogues increasingly difficult. Consequently, pre-training using self-supervised learning (SSL), which offers a scalable route by extracting structure directly from unlabelled images, is becoming attractive for many downstream applications. In this work we consider the use of coupled self-supervised representation learning approaches for radio galaxy morphology pre-training. In order to account for the more nuanced variations in radio galaxy morphology than are typically included in the augmented views of view-based SSL algorithms, we use a pre-trained Variational Autoencoder (VAE) to generate views for training a larger view-based self-supervised model. To do this, a $\beta$-VAE was trained on the Radio Galaxy Zoo (RGZ) dataset, where moderate regularization ($\beta = 2.3$) was found to provide a good balance between reconstruction quality and disentanglement of generative factors such as source multiplicity and lobe asymmetry. An analysis of the $\beta$-VAE reveals that Fanaroff-Riley class identity manifests as a continuous transition across the latent space, rather than being associated to a single discrete dimension. $\beta$-VAE reconstructions were then incorporated as generative augmentations within a view-based SSL pipeline. Our experiments show that combining these generative views with standard image augmentations improves downstream classification performance, and we present ablation studies clarifying the relative contribution of each augmentation type. These results indicate that generative and contrastive approaches are complementary, and point toward disentanglement-aware self-supervised learning as a promising direction for future radio astronomy surveys.
\end{abstract}

% Select between one and six entries from the list of approved keywords.
% Don't make up new ones.

%%%%%%%%%%%%%%%%%%%%%%%%%%%%%%%%%%%%%%%%%%%%%%%%%%

%%%%%%%%%%%%%%%%% BODY OF PAPER %%%%%%%%%%%%%%%%%%

\section{Introduction}

Radio galaxies, together with radio-loud quasars, are active galactic nuclei (AGN) distinguished by their strong emission at radio wavelengths. This emission is not confined to the central regions of the host galaxy but is often observed in the form of large-scale structures, including collimated jets and diffuse lobes. These jets can extend over distances of hundreds of kiloparsecs and, in extreme cases, up to several megaparsecs, making them among the largest single structures in the Universe. 

\cite{fanaroff1974} analysed 57 sources in the Third Cambridge Catalogue of Radio Sources (3CR) and introduced a morphological classification scheme based on the relative distribution of radio brightness. They defined the ratio between the separation of the brightest regions on opposite sides of the central galaxy and the total linear extent of the source. Sources with a ratio less than 0.5 were classified as Fanaroff–Riley type I (FR I), characterised by jets that are brightest near the galactic centre and fade with distance. Conversely, sources with a ratio greater than 0.5 were classified as Fanaroff–Riley type II (FR II), in which the outer edges dominate the emission and often display bright terminal hotspots.

While the Fanaroff–Riley scheme provides a binary framework, subsequent observations have revealed a much richer morphological diversity. Hybrid radio sources \citep[HyMoRS;][]{gopalkrishna2000extragalacticradiosourceshybrid} exhibit asymmetric structures, showing FR I-like characteristics on one side and FR II-like features on the other. These objects are extremely rare and are generally interpreted as the result of environmental asymmetries affecting the propagation of relativistic jets. Narrow-angle tail (NAT) and wide-angle tail (WAT) sources, also referred to collectively as bent-tail galaxies \citep{headtail}, are associated with dense cluster environments. Their characteristic bent jets are thought to arise from the relative motion of the host galaxy through the intracluster medium, providing an important diagnostic of environmental effects. Compact radio galaxies, sometimes termed FR 0s, represent another significant class. These sources lack large-scale jets or lobes and instead display emission concentrated near the nucleus \citep{Baldi_2015}. They are thought to represent either young radio galaxies whose jets have yet to extend far from the central engine or systems in which the jet activity is disrupted or confined. Other exotic morphologies, such as X-shaped or double-double radio galaxies, further illustrate the variety of forms that can emerge from the interaction of jets, host galaxies, and surrounding environments. 

Classifying radio galaxies becomes increasingly challenging as observations extend to fainter sources and more distant populations. At low signal-to-noise ratios or with limited resolution, it is often difficult to distinguish morphological features clearly, and many sources show intermediate structures that do not fit neatly into simple categories. Building large, statistically meaningful samples across different redshifts, flux ranges, and environments is therefore essential for understanding radio galaxy evolution. However, the vast amount of data produced by modern surveys makes traditional visual, “by-eye” classification methods impractical.

Recent surveys highlight the scale of this challenge. The LOFAR Two-metre Sky Survey (LoTSS) Data Release 2, conducted with the LOw-Frequency ARray (LOFAR), has generated approximately 7 petabytes of data containing nearly 4.4 million sources \citep{Shimwell_2022}. The Very Large Array Sky Survey (VLASS) is expected to detect around 5.3 million sources, with the first epoch alone cataloguing 1.9 million \citep{Lacy_2020, Gordon_2021}. The Rapid ASKAP Continuum Survey (RACS) with the Australian SKA Pathfinder (ASKAP) has identified roughly 2.6 million sources in its high-frequency band (RACS-High, 1655.5 MHz;  \citealt{duchesne2025rapidaskapcontinuumsurvey}). The GaLactic and Extragalactic All-sky MWA eXtended (GLEAM-X) survey, conducted with the Murchison Widefield Array (MWA), contains about 625,000 sources \citep{ross2024galacticextragalacticallskymurchison}. Similarly, the MeerKAT International GHz Tiered Extragalactic Exploration Survey (MIGHTEE) has produced approximately 144,000 sources from its lower-resolution images and 114,000 from higher-resolution data \citep{hale2024mighteecontinuumsurveydata}. Earlier large-scale surveys, such as the Faint Images of the Radio Sky at Twenty Centimeters (FIRST; \citealt{first1995}) and the NRAO VLA Sky Survey (NVSS; \citealt{nvss1998}), together catalogued on the order of 580,000 sources over their operational lifetimes \citep{Kimball_2013}.

With the accelerating pace of data acquisition, manual classification is no longer a sustainable approach. Automated methods, ranging from classical statistical techniques to modern machine learning algorithms, have therefore become essential for processing and interpreting the growing wealth of radio data. These approaches aim not only to reproduce traditional morphological categories such as FR I and FR II, but also to uncover new patterns and subclasses that emerge when millions of sources are analysed in a uniform and scalable way.

\subsection{Machine Learning in Radio Galaxy Classification}

Although machine learning has enabled significant progress in the classification and representation of radio galaxies {\citep[see e.g.][]{Aniyan_2017, Alhassan_2018, tang2019, mohan2024evaluatingbayesiandeeplearning}, several challenges remain that limit the effectiveness of current approaches. A central constraint is the scarcity of large, balanced, and consistently labelled datasets. Catalogues such as MiraBest \citep{mirabestdataset} and subsets of Radio Galaxy Zoo \citep{Banfield_2015} provide valuable training resources, but they suffer from class imbalance, with FRI sources underrepresented relative to FRII galaxies, and rare morphologies such as hybrids or bent-tailed sources appearing only sparsely. Furthermore, reliable labelling of morphology requires expert input and is susceptible to inconsistencies in large community-driven projects. These limitations restrict the performance of supervised methods, which depend on abundant, high-quality labels. At the same time, upcoming surveys such as the SKA are expected to produce data on an unprecedented scale, intensifying the need for methods that can generalize beyond small, carefully annotated samples.

As supervised learning models rely heavily on labelled data, many recent studies have investigated semi-supervised and unsupervised methods that can leverage large unlabelled datasets. \cite{Baron_Perez_2025} applied deep clustering with semi-supervised learning to classify LoTSS-DR2 sources into multiple morphological categories, showing that unlabelled data could significantly improve classification accuracy. Unsupervised approaches based on self-organizing maps (SOMs) and autoencoders have also been used to discover clusters of radio morphologies without human supervision (\citealt{ralph2019, Mostert_2021}). These methods are particularly relevant because the morphological diversity of radio galaxies extends beyond the canonical FRI/FRII dichotomy, encompassing classes such as hybrids, X-shaped, and double-double sources. By grouping sources in an unsupervised way, these approaches allow astronomers to explore unexpected structures and potentially identify new categories of radio galaxies. 

A further challenge is that standard machine learning methods capture only a limited range of morphological diversity. Common image augmentations such as flips, rotations, crops, and intensity changes help models generalize but are not connected to the underlying astrophysics. They cannot reproduce variations tied to physical properties of the source or its environment, such as changes in jet length, lobe curvature, brightness asymmetry, or core prominence. Simulations of radio galaxies, created using hydrodynamical or semi-empirical models, can generate such physically meaningful diversity \citep{bonaldi2018ska, Turner_2022}. However, these simulations are computationally expensive and not yet feasible at the scale required by surveys like the SKA. This creates a gap between the need for morphologically diverse training data and the limitations of both labelled catalogues and physical simulations.

Generative models offer a complementary approach that addresses these limitations from a data-driven perspective. These models can learn compressed latent representations of radio galaxies and can generate new, plausible variations of sources that interpolate between and extend beyond the training examples, effectively creating "new views" of radio galaxies. Such representations surpass simple image manipulations by reflecting physical properties of the sources, for example, differences in jet structure, lobe asymmetry, or the prominence of the central core. Unlike traditional augmentations, which operate through predefined geometric transformations, generative augmentations have the potential to introduce variations that are more closely aligned with the intrinsic diversity of the data itself. This makes them a natural candidate for integration into self-supervised learning pipelines, where the goal is to learn robust and generalizable representations without relying on large amounts of labelled data.

This strategy parallels methods that train on synthetic data derived from hydrodynamical or semi-empirical simulations \citep{10.1111/j.1365-2966.2008.13486.x, bonaldi2018ska, Turner_2022}. However, instead of relying on external, computationally intensive simulations to expand the training set, our approach integrates the generative process directly into the self-supervised pipeline. By learning the underlying generative factors of the data distribution directly \citep{Higgins2016betaVAELB}, the model produces 'internal simulations' (new views) that are statistically consistent with the observed population. This allows for a self-contained framework where the augmentation strategy evolves alongside the representation learner.

The motivation of this work is therefore twofold. First, to critically evaluate whether generative views can contribute to self-supervised representation learning in a way that is competitive with, or complementary to, conventional augmentations. Second, to explore how disentangled latent representations might provide interpretable insights into the structure of radio galaxy morphologies.

\section{Self-supervised Learning}

Traditional machine learning methods, such as supervised learning, depend on large, labelled datasets in which models are trained to associate inputs with known outputs. Although highly successful in many applications, this approach is less practical in domains such as astronomy, where data volumes are vast but reliable labels are scarce and require expert knowledge. Self-supervised learning (SSL) addresses this challenge by exploiting the intrinsic structure of unlabelled data to formulate pretext tasks, allowing models to learn informative representations without direct supervision \citep{ssljing}. These learned representations can then be adapted to downstream tasks such as classification or clustering, thereby reducing reliance on labelled data.

A key motivation for SSL is the problem of dimensionality reduction. Astronomical images, such as those of radio galaxies, are inherently high-dimensional, with thousands of pixels encoding complex and often subtle structures. To enable effective analysis, this information must be compressed into compact latent representations that retain the relevant features while discarding noise and redundancy. Early approaches to this problem relied on linear techniques such as principal component analysis (PCA; \citealt{shlens2014pca}), but these methods are limited in their ability to capture nonlinear relationships. Neural network–based methods, such as autoencoders, have since become dominant due to their ability to capture nonlinear structures in the data. SSL builds upon this idea by defining training objectives that force networks to extract robust, semantically meaningful features directly from raw inputs.

Two major paradigms of SSL have emerged in recent years. The first is reconstruction-based (generative) SSL, where models such as autoencoders learn to compress and reconstruct inputs, sometimes with modifications such as masked regions that must be predicted \citep{he2021maskedautoencodersscalablevision}. The second is view-based SSL, where models are trained to produce consistent representations of different augmented “views” of the same input, using either contrastive \citep{chen2020simclr} or non-contrastive frameworks \citep{grill2020byol}. Both approaches have shown state-of-the-art performance across computer vision and natural language processing, and are increasingly being explored in astronomy \citep{Hayat_2021, Marianer_2020, slijepcevic2023radiogalaxyzoobuilding, lastufka2025bridginggapexaminingvision, Parker_2024}.

\subsection{View-based SSL}

In view-based learning, models are trained to produce consistent representations of different “views” of the same input. Unlike reconstruction-based approaches, which aim to recover missing or corrupted input data, view-based SSL focuses on learning invariance: the representation should remain stable under transformations such as cropping, rotation, color distortion, or noise \citep{Le_Khac_2020, ssljing}. This principle is particularly valuable in domains such as astronomy, where sources may appear at different scales, orientations, or noise levels depending on the instrument and observing conditions.

The key idea is to generate two or more augmented versions of an input image and then train the network so that their latent representations are close to one another. This forces the model to capture underlying content rather than the pixel-level differences. Approaches to view-based SSL can be divided into two categories: contrastive and non-contrastive.

In astronomy, view-based self-supervised learning is particularly appealing because augmentations can be tailored to observational data. For example, rotated or mirrored versions of radio galaxy images remain morphologically valid, while noise injection can mimic instrumental effects. By enforcing invariance across such transformations, view-based SSL enables the discovery of representations that capture underlying astrophysical structure rather than imaging artifacts \citep{cecconello2024selfsupervisedlearningradioastronomysource}.

One widely used framework in view-based SSL is SimCLR \citep{chen2020simclr}, which learns representations by encouraging two augmented views of the same image to be mapped to nearby points in the latent space, while ensuring that views from different images remain well separated. This is achieved through a \emph{contrastive} loss function, which quantifies similarity between representations using measures such as cosine similarity, and regulates the sharpness of the separation with a temperature parameter. In astronomy, contrastive methods have been explored for learning galaxy morphology representations \citep[e.g.][]{Hayat_2021}. However, the reliance of contrastive methods on large numbers of negative pairs can typically result in large batches or memory banks to provide sufficient negative pairs, which can be computationally expensive and difficult to scale. To address this limitation, non-contrastive self-supervised methods were introduced. These approaches eliminate the need for explicit negative examples and instead use carefully designed architectures or optimization strategies to prevent the trivial solution of representational collapse, where the network outputs the same representation for every input, making the features useless.

One influential approach is Bootstrap Your Own Latent (BYOL; \citealt{grill2020byol}), which introduced asymmetric online and target networks updated through a momentum encoder. BYOL demonstrated that non-contrastive methods could achieve performance competitive with, or even exceeding, contrastive methods on standard benchmarks.

\section{BYOL: Bootstrap Your Own Latent}

Bootstrap Your Own Latent \citep[BYOL;][]{grill2020byol} is a non-contrastive learning method that represents a significant step in the development of view-based SSL. Unlike contrastive frameworks, which explicitly require negative pairs, BYOL learns representations by aligning two augmented views of the same image using only positive pairs. The key insight of BYOL is that collapse is avoided through an asymmetric architecture consisting of an online network and a target network, coupled with a stop-gradient mechanism.

The BYOL framework operates on two stochastic augmentations of an input image, 
\(\tilde{\mathbf{x}}_1\) and \(\tilde{\mathbf{x}}_2\). Each augmentation is passed through an encoder followed by a projection head to obtain representations. The online network 
consists of an encoder \(f_\theta\), a projection head \(g_\theta\), and an additional predictor \(p_\theta\). 
The target network, on the other hand, consists only of an encoder \(f_\xi\) and projection head \(g_\xi\).  

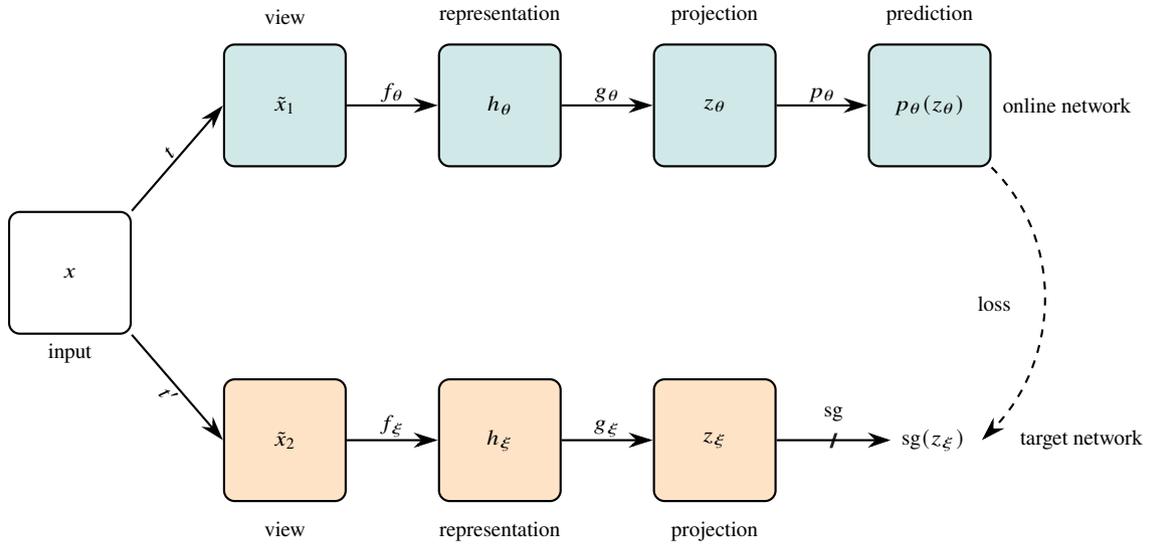
\begin{figure*}
    % \centering
\begin{tikzpicture}[
  box/.style={draw, rounded corners, minimum size=1.6cm, align=center, thick},
  online/.style={box, fill=teal!18},
  target/.style={box, fill=orange!22},
  lab/.style={font=\footnotesize, inner sep=1pt},
  arr/.style={-{Stealth[length=3.2mm,width=2mm]}, thick}
]

% ---------- Input ----------
\node[box, minimum size=1.6cm] (xin) at (0,0) {$x$};
\node[lab, below=2pt of xin.south] {input};

% ---------- ONLINE (top row) ----------
\node[online, right=1.2cm of xin, yshift=2.2cm] (x1) {$\tilde{x}_1$};
\node[online, right=1.2cm of x1] (h)  {$h_\theta$};
\node[online, right=1.2cm of h]  (z)  {$z_\theta$};
\node[online, right=1.2cm of z]  (p)  {$p_\theta(z_\theta)$};

% Stage labels (top)
\node[lab, above=6pt of x1] {view};
\node[lab, above=6pt of h]  {representation};
\node[lab, above=6pt of z]  {projection};
\node[lab, above=6pt of p]  {prediction};

% Online arrows
\draw[arr] (xin.north east) -- (x1.west) node[lab, midway, sloped, above] {$t$};
\draw[arr] (x1.east) -- (h.west) node[lab, midway, above] {$f_\theta$};
\draw[arr] (h.east)  -- (z.west) node[lab, midway, above] {$g_\theta$};
\draw[arr] (z.east)  -- (p.west) node[lab, midway, above] {$p_\theta$};

% Online label
\node[lab, right=3pt of p.east] {online network};

% ---------- TARGET (bottom row) ----------
\node[target, right=1.2cm of xin, yshift=-2.2cm] (x2) {$\tilde{x}_2$};
\node[target, right=1.2cm of x2] (ht) {$h_\xi$};
\node[target, right=1.2cm of ht] (zt) {$z_\xi$};

% Stage labels (bottom)
\node[lab, below=6pt of x2] {view};
\node[lab, below=6pt of ht] {representation};
\node[lab, below=6pt of zt] {projection};

% Target arrows
\draw[arr] (xin.south east) -- (x2.west) node[lab, midway, sloped, below] {$t'$};
\draw[arr] (x2.east) -- (ht.west) node[lab, midway, above] {$f_\xi$};
\draw[arr] (ht.east) -- (zt.west) node[lab, midway, above] {$g_\xi$};

% Stop-gradient stub (with // slash)
\coordinate (sgstart) at (zt.east);
\coordinate (sgend) at ($(zt.east)+(1.5,0)$);
\draw[-{Stealth[length=3mm,width=2mm]}, thick] (sgstart) -- (sgend);
\node[lab, right=3pt of sgend] {$\mathrm{sg}(z_\xi)$};
% Double slash marker in the middle
\node[lab] (slash) at ($(sgstart)!0.5!(sgend)$) {/\!\!/};
% "sg" label above the marker
\node[lab, above=2pt of slash] {sg};
% Dashed loss arrow from predictor to stop-gradient output
\draw[arr, dashed] (p.south east) .. controls +(1.0,-1.0) and +(1.0,1.0) .. ($(sgend)+(1.2,0)$);
\node[lab] at ($(p.east)!0.5!(sgend)+(0.7,-0.4)$) {loss};

% Target label
\node[lab, right=90pt of zt.east] {target network};

\end{tikzpicture}
    \caption{Schematic of the BYOL framework with online and target networks.}
    \label{fig:byol}
\end{figure*}

Given an augmented input \(\tilde{\mathbf{x}}_1\), the online network computes  
\begin{equation}
\mathbf{z}_1 = g_\theta(f_\theta(\tilde{\mathbf{x}}_1)), 
\quad 
\mathbf{p}_1 = p_\theta(\mathbf{z}_1), 
\end{equation}
and the target network processes \(\tilde{\mathbf{x}}_2\) as  
\begin{equation}
\mathbf{z}_2 = g_\xi(f_\xi(\tilde{\mathbf{x}}_2)).
\end{equation}

The online prediction \(\mathbf{p}_1\) is trained to match the target representation \(\mathbf{z}_2\). 
Symmetrically, the roles of the two augmentations are swapped, yielding a bidirectional 
training objective.

\subsection{Training Objective}

The training objective in BYOL is to align the representations produced by the online and target networks for two augmented views of the same input. This is achieved by minimizing the distance between the online predictor’s output and the target projection, with similarity measured using a cosine-based loss, which is given by, 
\begin{equation}
\mathcal{L}_{\text{BYOL}} = \| \mathbf{q}_1 - \text{sg}(\mathbf{z}_2) \|_2^2 + \| \mathbf{q}_2 - \text{sg}(\mathbf{z}_1) \|_2^2,
\end{equation}
where $q_1$ and $q_2$ are the online predictions, $z_1$ and $z_2$ are the target projections, and $sg()$ denotes the stop-gradient operator. The stop-gradient prevents the target network from receiving gradient updates, ensuring that optimization only affects the online network. Without this asymmetry, the two networks would converge to a solution in which all representations collapse to a constant vector.

The parameters of the target network are not learned directly through backpropagation. Instead, they are updated as an exponential moving average (EMA) of the online network’s parameters:
\begin{equation}
\xi \leftarrow \tau \xi + (1 - \tau)\theta,
\end{equation}
where $\theta$ are the online parameters, $\xi$ are the target parameters, and $\tau$ is a momentum coefficient. This slow updating mechanism stabilizes training by providing a steadily evolving target representation, against which the online network can consistently align.

%\subsection{BYOL in Astronomy}

%Astronomical surveys generate vast collections of images, yet only a small fraction of these data are accompanied by reliable labels. BYOL is well suited to this setting, as it can exploit large collections of unlabelled observations to learn informative representations. Through augmentations such as rotations, cropping, flipping, and noise perturbations, BYOL can capture features that remain stable across observational variations while still encoding essential morphological information.

BYOL avoids the need for large batches or memory banks, which are common requirements in contrastive methods. This makes it computationally more efficient and adaptable to research environments where resources are constrained. For astronomy, this efficiency is especially valuable given the scale of modern surveys. Early applications of SSL in galaxy morphology analysis \citep{Hayat_2021} demonstrate the feasibility of such approaches, and BYOL offers a strong framework for building robust representations of radio galaxy morphologies \citep[e.g.][]{slijepcevic2023radiogalaxyzoobuilding}.

\section{Variational Autoencoders (VAEs)}
Variational Autoencoders (VAEs) are a class of deep generative models that combine neural network–based autoencoders with the principles of variational inference \citep{kingma2022autoencodingvariationalbayes}. Unlike classical autoencoders, which learn a deterministic mapping between inputs and compressed latent representations, VAEs impose a probabilistic structure on the latent space by approximating intractable posteriors with variational distributions. 

The VAEs have a dual role as they act as both representation learners, compressing high-dimensional data into meaningful latent variables, and as generative models, capable of producing synthetic data that reflects the statistical properties of the training set. These capabilities make VAEs very valuable in astronomy, where they can be used for data augmentation and disentanglement of physical features.

The architecture of a VAE consists of two neural networks: an encoder (the inference network) and a decoder (the generative network). The encoder approximates the posterior distribution over latent variables, while the decoder reconstructs or generates data from samples drawn from those variables.

A prior distribution is placed over the latent variables, usually an isotropic Gaussian,
\begin{equation}
    p(z) = \mathcal{N}(z; 0,I),
\end{equation}
which regularizes the latent space and enables straightforward sampling.

The encoder then maps an input $x$ and produces the parameters of an approximate posterior distribution over the latent variables $z$. Following \cite{kingma2022autoencodingvariationalbayes}, this approximate posterior is modeled as a multivariate Gaussian with a diagonal covariance matrix:
\begin{equation}
    \log q_\phi({z} \mid {x}) = \log \mathcal{N}\!\left({z}; \, {\mu}_{\phi}, \,{\sigma}_{\phi}^2 I\right),
    \label{eqn : multgauss}
\end{equation}
where ${\mu}_{\phi}$ and  ${\sigma}_{\phi}^2$ are the mean and variance vector, predicted by the encoder with parameters $\phi$.

The decoder, parameterized by $\theta$, defines the likelihood,
$p_{\theta}(x \mid z)$,
which reconstructs the input (or generates new samples) from latent codes.

The key element of the VAE is the reparameterization trick, which makes sampling differentiable. Instead of sampling $z$ directly from the encoder’s distribution, it is expressed as
\begin{equation}
    z = \mu_{\phi}(x) + \sigma_{\phi}(x) \odot \epsilon, \quad \epsilon \sim \mathcal{N}(0,I)
\end{equation}
where $\odot$ denotes elementwise multiplication. This allows gradients to flow through the encoder during training, enabling end-to-end optimization with stochastic gradient descent.

\subsection{Training Objective}

Training a VAE requires optimizing a loss function that balances two competing goals. The first is to accurately reconstruct the input data, and the second is to ensure that the latent space remains structured according to a prior distribution. This objective is formalized through the Evidence Lower Bound (ELBO), which arises from variational inference.
The true posterior, $p_\theta(z \mid x)$, is analytically intractable because the marginal likelihood, $p_\theta (x)$, cannot be computed exactly for high dimensional latent spaces. Variational inference (VI) provides a tractable solution by introducing an approximate posterior $q_\phi(z \mid x)$.

\subsubsection{The Evidence Lower Bound (ELBO)}
The central idea is to maximize the log marginal likelihood of the data, 
$\log p_\theta(x)$. Since this is intractable, we derive a 
variational lower bound using Jensen’s inequality 
\citep{jordan1999, Blei03042017},
\begin{eqnarray}
\log p_\theta({x}) 
&=& \log \int q_\phi(z \mid x) 
    \frac{p_\theta(x, z)}{q_\phi(z \mid x)} \, dz \\
&\geq& \, \mathbb{E}_{q_\phi(z \mid x)} 
    \big[ \log p_\theta(x, z) - \log q_\phi(z \mid x) \big].
\end{eqnarray}
This inequality defines the Evidence Lower Bound (ELBO),
\begin{equation}
\mathcal{L}(\theta, \phi; x) 
= \mathbb{E}_{q_\phi(z \mid x)} 
    \big[ \log p_\theta(x \mid z) \big] 
    - D_{\text{KL}} \big( q_\phi(z \mid x) \,\|\, p_\theta{(z)} \big).
\end{equation}
Maximizing the ELBO simultaneously encourages two properties,
\begin{enumerate}
    \item The first term ensures that the decoder reconstructs the observed data accurately from the latent variables.
    \item The second term regularizes the approximate posterior to remain close to the prior \(p_\theta{(z)}\), typically chosen as an isotropic Gaussian.
\end{enumerate}

Thus, the ELBO provides a balance between data reconstruction and latent space regularization.

\subsubsection{Reconstruction Loss}
The reconstruction loss, acting as the data fidelity term,
\begin{equation}
    \mathbb{E}_{q_\phi(z \mid x)} 
    \big[ \log p_\theta(x \mid z) \big] 
\end{equation}
encourages the decoder to produce outputs that closely match the observed data when conditioned on the latent variables. In practical implementations, this expectation reduces to a likelihood-based loss function that depends on the choice of observation model. For continuous valued data, the likelihood is typically modeled as a Gaussian distribution, which corresponds to an MSE loss. For binary-valued or normalized pixel data, a Bernoulli likelihood is commonly adopted, yielding a cross-entropy loss.

This component of the objective ensures that the latent representation retains sufficient information for accurate reconstruction, effectively measuring how well the VAE compresses and regenerates the input data.

\subsubsection{KL Divergence}
The second component of the ELBO corresponds to the Kullback–Leibler (KL) divergence, which measures how far the approximate posterior deviates from the prior distribution, given by,
\begin{equation}
    D_{\text{KL}} \big( q_\phi(z \mid x) \,\|\, p_\theta{(z)} \big)
\end{equation}

In the standard VAE formulation, the approximate posterior is modelled as a Gaussian distribution with diagonal covariance (Equation~\ref{eqn : multgauss}) while the prior is typically taken to be $p_\theta{(z)} = \mathcal{N}(z; 0,I).$

Given these choices, the KL divergence can be simplified to a closed-form expression,
\begin{equation}
D_{\text{KL}}\big(q_\phi(z \mid x) \,\|\, p_\theta{(z)}\big) 
= \frac{1}{2} \sum_{i=1}^{d} \left( \sigma_{\phi,i}^2 + \mu_{\phi,i}^2 - 1 - \log \sigma_{\phi,i}^2 \right),
\end{equation}
where \(d\) denotes the dimensionality of the latent space.

This regularization term enforces that the approximate posterior does not drift too far from the prior. As a result, the latent space remains continuous, compact, and well-behaved, which is crucial for meaningful sampling and interpolation in generative tasks.

\subsection{$\beta$-VAE and Disentanglement}

A central challenge in representation learning is to capture underlying generative factors of variation in data in a way that is both interpretable and useful for downstream tasks. Standard VAEs often entangle these factors in the latent space, making it difficult to separate distinct attributes such as shape, orientation, or brightness in image data. To address this limitation, the $\beta$-VAE was introduced as a modification of the original VAE framework to promote ``disentangled'' representations \citep{Higgins2016betaVAELB}.

The key idea of a $\beta$-VAE is to alter the standard VAE objective by scaling the KL divergence term in the ELBO with a hyperparameter $\beta$,
\begin{align}
\nonumber \mathcal{L}_{\beta-VAE}(\theta, \phi; x) 
= \mathbb{E}_{q_\phi(z \mid x)} &
    \big[ \log p_\theta(x \mid z) \big] \\
    & - \beta \,D_{\text{KL}} \big( q_\phi(z \mid x) \,\|\, p_\theta{(z)} \big).
\end{align}

When $\beta=1$, the formulation reduces to the standard VAE objective. By setting $\beta>1$, the model places greater emphasis on aligning the approximate posterior with the prior distribution, effectively constraining the capacity of the latent space. This constraint reduces the capacity of the latent space, encouraging the model to represent distinct generative factors independently across different latent dimensions. A common training strategy is to apply KL annealing, where the $\beta$ term is gradually increased from zero to its target value, mitigating posterior collapse and enabling the model to learn useful representations before enforcing strong regularization \citep{bowman2016generatingsentencescontinuousspace}.

$\beta$-VAE demonstrated for the first time that disentangled representations could emerge in an unsupervised setting, without requiring labelled factors of variation \citep{Higgins2016betaVAELB, burgess2018understandingdisentanglingbetavae}. This was a key conceptual advance, showing that structural constraints on probabilistic models can yield interpretable representations aligned with the independent causal mechanisms underlying the data. However, $\beta$-VAE introduces a trade-off between disentanglement and reconstruction quality; higher values of $\beta$ encourage greater disentanglement but often degrade the fidelity of reconstructions.

To quantify disentanglement, \citet{Higgins2016betaVAELB} proposed one of the earliest metrics designed to evaluate whether variations in generative factors are captured by individual latent dimensions. While this metric has been influential, it is limited by its reliance on access to ground-truth generative factors and has been criticized for instability and weak correlation with downstream task performance \citep{locatello2019challengingcommonassumptionsunsupervised}. Consequently, latent traversals remain a widely used alternative: by systematically varying one latent variable while keeping others fixed, researchers can directly observe whether interpretable factors of variation emerge. This approach, applied in both early and subsequent studies \citep{Higgins2016betaVAELB, burgess2018understandingdisentanglingbetavae, locatello2019challengingcommonassumptionsunsupervised, zhu2020learningdisentangledrepresentationslatent}, continues to provide intuitive and compelling evidence of disentanglement in practice.

\subsection{Other Generative Models}

Beyond VAEs, several other generative frameworks have become central to modern machine learning. Generative Adversarial Networks (GANs) \citep{goodfellow2014generativeadversarialnetworks} train a generator and discriminator in an adversarial game, producing highly realistic samples but often facing instability and mode collapse. Normalizing Flows \citep{rezende2016variationalinferencenormalizingflows} model complex distributions through invertible transformations of simple priors, allowing exact likelihood computation and flexible density estimation, though scalability can be restrictive. Diffusion Models \citep{sohldickstein2015deepunsupervisedlearningusing} generate data by reversing a noise corruption process. They achieve state-of-the-art synthesis quality and diversity, albeit at high computational cost due to iterative sampling.

Together, these approaches complement VAEs by offering different trade-offs between sample quality, tractability, and efficiency, shaping the landscape of deep generative modeling.

\subsection{Source Simulation in Astronomy}

Simulating astronomical sources has become an essential tool for both data analysis and methodological development in modern astronomy. As next-generation surveys like the SKA generate unprecedented volumes of data, simulations provide a controlled environment to test detection pipelines, evaluate classification algorithms, and understand systematic biases. Synthetic sources allow researchers to benchmark machine learning models where labelled data are scarce, and to explore parameter spaces that are underrepresented in current observations.

In radio astronomy, simulation frameworks have been developed to model extragalactic radio sources across a range of morphologies and physical conditions. For example, the SKADS Simulated Sky (S³) project \citep{10.1111/j.1365-2966.2008.13486.x} generated large-scale simulations of radio galaxies, star-forming galaxies, and AGN populations to support survey planning and cosmological studies. More recently, the Tiered Radio Extragalactic Continuum Simulation \citep[T-RECS;][]{Bonaldi_2018trecs} has been introduced, providing improved modeling of AGN and star-forming populations, including their spectral, polarization, and clustering properties. Such synthetic datasets are invaluable for probing the limitations of observational sensitivity and resolution.

Alongside such astrophysical simulations, generative models are increasingly being employed to produce realistic source populations in a data-driven manner. In the optical domain, the Celeste model applied variational inference to construct a probabilistic generative framework of stars and galaxies, explicitly incorporating observational effects such as noise and point spread functions \citep{regier2015celestevariationalinferencegenerative}. In radio astronomy, structured variational autoencoders have been used to synthesize FR I and FR II galaxies, generating morphology-consistent images that aid both survey preparation and the development of machine learning pipelines \citep{Bastien_2021} and more recently diffusion models have been developed for the same purpose \citep{martinez2024}.

\section{Datasets}
The experiments in this thesis are conducted on two benchmark datasets of radio galaxies: the MiraBest dataset \citep{porter2023mirabestdatasetmorphologicallyclassified} and the Radio Galaxy Zoo (RGZ) catalogue \citep{Banfield_2015}. Both datasets contain examples of FR type I and II morphologies, but they differ considerably in scale, curation, and labelling methodology, making them complementary resources for studying machine learning approaches in radio astronomy.

The MiraBest dataset is a carefully curated sample of FR I and FR II sources with expert-confirmed classifications. Its relatively modest size is offset by the high reliability of its labels, which makes it particularly well-suited as a benchmark for evaluating model performance and comparing experimental outcomes.

By contrast, the RGZ dataset represents one of the largest citizen-science efforts in astronomy, in which thousands of volunteers visually matched radio emission to host galaxies. While RGZ does not directly provide FR class labels, it captures a far broader and more diverse population of radio sources. With appropriate filtering or the application of automated reclassification pipelines, it can be leveraged as a large-scale training and testing ground for machine learning methods.

Together, these datasets serve complementary roles in this work. RGZ, with its large scale and morphological diversity, is used to train both the $\beta$-VAE and the BYOL model in a fully unsupervised manner. MiraBest, with its expert labels, is used exclusively for downstream evaluation, training the linear classifier and fine-tuning the encoder to assess the quality of the learned representations.

\subsection{MiraBest dataset}
The MiraBest dataset \citep{porter2023mirabestdatasetmorphologicallyclassified} is a curated collection of 1,256 radio-loud AGN drawn from the work of \cite{Miraghaei2017TheEnvironment}, who cross-matched the Sloan Digital Sky Survey (SDSS DR7; \citealt{Abazajian_2009}) with radio sources from NVSS \citep{nvss1998} and FIRST \citep{first1995}. Each source was visually inspected and assigned a FR morphological class as either FRI or FRII. In addition to these FR types, the dataset also contains hybrid sources, as well as subclass labels for non-standard morphologies such as double-double, WATs, diffuse, and head-tail structures.

Each source is encoded with a three-digit identifier (Table \ref{tab:mirabest_digits}). The first digit indicates the FR class, the second digit encodes classification confidence, and the third digit specifies the morphological subclass.

\begin{table}
\centering
\caption{Three digit identification scheme for sources in \citet{Miraghaei2017TheEnvironment}.}
\renewcommand{\arraystretch}{1.3} % row spacing
\setlength{\tabcolsep}{12pt} % column spacing
\begin{tabular}{l l l}
\hline
Digit 1 & Digit 2 & Digit 3 \\
\hline
1: FRI           & 0: Confident  & 0: Standard \\
2: FRII          & 1: Uncertain  & 1: Double Double \\
3: Hybrid        &               & 2: Wide Angle Tail \\
4: Unclassifiable&               & 3: Diffuse \\
                 &               & 4: Head Tail \\
\hline
\end{tabular}
\label{tab:mirabest_digits}
\end{table}

The overall composition of the dataset is summarised in Table \ref{tab:mirabest_class}. There are 591 FR I sources, 631 FR II sources, and 34 hybrids, with both confident and uncertain examples. The FR I and FR II populations are relatively balanced, which is particularly advantageous for machine learning applications, as these two classes form the canonical FR morphologies and serve as a benchmark for testing classification and representation learning approaches. The inclusion of hybrid sources, while comparatively rare, is also valuable since such objects exhibit mixed or transitional features between the two canonical classes and therefore provide an opportunity to study edge cases that challenge traditional classification schemes.

\begin{table}
\centering
\caption{Population of sources in the MiraBest dataset}
\renewcommand{\arraystretch}{1.2} % row spacing
\setlength{\tabcolsep}{12pt} % column spacing
\begin{tabular}{l c c}
\hline
Class & Confidence & No. \\
\hline
FRI   & Confident  & 397 \\
      & Uncertain  & 194 \\
FRII  & Confident  & 436 \\
      & Uncertain  & 195 \\
Hybrid& Confident  & 19  \\
      & Uncertain  & 15  \\
\hline
\end{tabular}
\label{tab:mirabest_class}
\end{table}

To make the dataset compatible with machine learning workflows, preprocessing steps were applied following the pipeline of \cite{tang2019transfer}. For each source, a 300×300 pixel FIRST image cutout was downloaded, centered on the radio source position. Images were then cropped to 150×150 pixels, sigma-clipped to suppress background noise, and circularly masked to remove unrelated sources in the field. Finally, pixel intensities were normalized to the range [0, 255] and converted to PNG format for efficient storage and loading. Example cutouts from the MiraBest dataset are shown in Figure \ref{fig:three_images}, illustrating confident samples of FR I, FR II, and Hybrid morphologies.

This combination of high-quality expert labels, confidence annotations, and well-documented preprocessing makes MiraBest an ideal dataset for benchmarking generative modeling and self-supervised learning methods in radio astronomy.

\begin{figure*}
\centerline{(a) Confident FR~I}
\centerline{\includegraphics[width=0.95\textwidth]{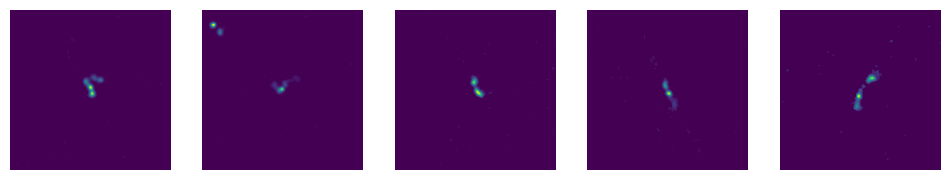}} 
\centerline{(b) Confident FR~II}
\centerline{\includegraphics[width=0.95\textwidth]{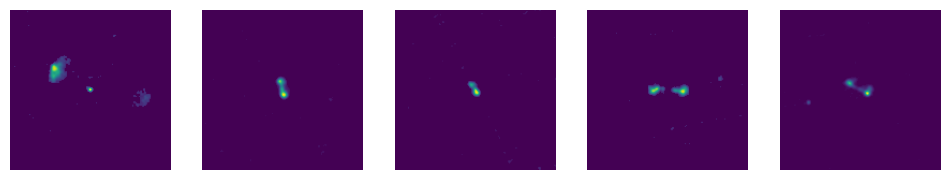}} 
\centerline{(c) Confident Hybrid}
\centerline{\includegraphics[width=0.95\textwidth]{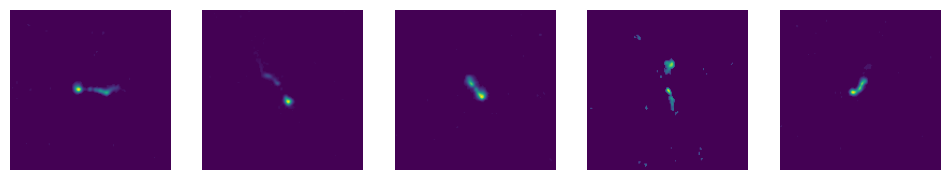}}
\caption{Example radio galaxy cutouts from the MiraBest dataset, showing confident samples of (a) FR I, (b) FR II, and (c) Hybrid morphologies.}
    \label{fig:three_images}
\end{figure*}

\subsection{Radio Galaxy Zoo (RGZ)}
The RGZ project is a large-scale citizen science initiative aimed to address the challenge of cross-identifying radio sources with their corresponding host galaxies \citep{Banfield_2015}. Traditional automated methods often fail for extended or morphologically complex radio sources, particularly those with multiple lobes, jets, or irregular structures. RGZ leverages the collective effort of thousands of volunteers to perform visual classifications, thereby enabling the construction of large and accurate training datasets for future machine learning applications in radio astronomy.

The dataset primarily draws from two deep 1.4 GHz radio continuum surveys: the FIRST \citep{first1995} and the Australia Telescope Large Area Survey (ATLAS; \citealt{atlas2015}), complemented with mid-infrared imaging from the Wide-field Infrared Survey Explorer (WISE; \citealt{Wright_2010}) and the Spitzer Wide-area Infrared Extragalactic Survey (SWIRE; \citealt{Lonsdale_2003}). Volunteers classify radio components within $3' \times 3'$ fields and identify their most likely infrared counterparts, producing consensus-based host galaxy associations. In the first year of operation, RGZ collected over one million classifications, covering $\sim 175,000$ radio sources \citep{Banfield_2015}. Example cutouts from the RGZ dataset are shown in Figure \ref{fig:rgz_cutouts}.

The classifications are probabilistic and consensus-driven rather than absolute, making the dataset effectively unlabelled in a morphological sense. Unlike curated datasets such as MiraBest, RGZ does not provide definitive FR classes but instead facilitates host identification and morphology clustering. 

\begin{figure*}
    \includegraphics[width=0.95\textwidth]{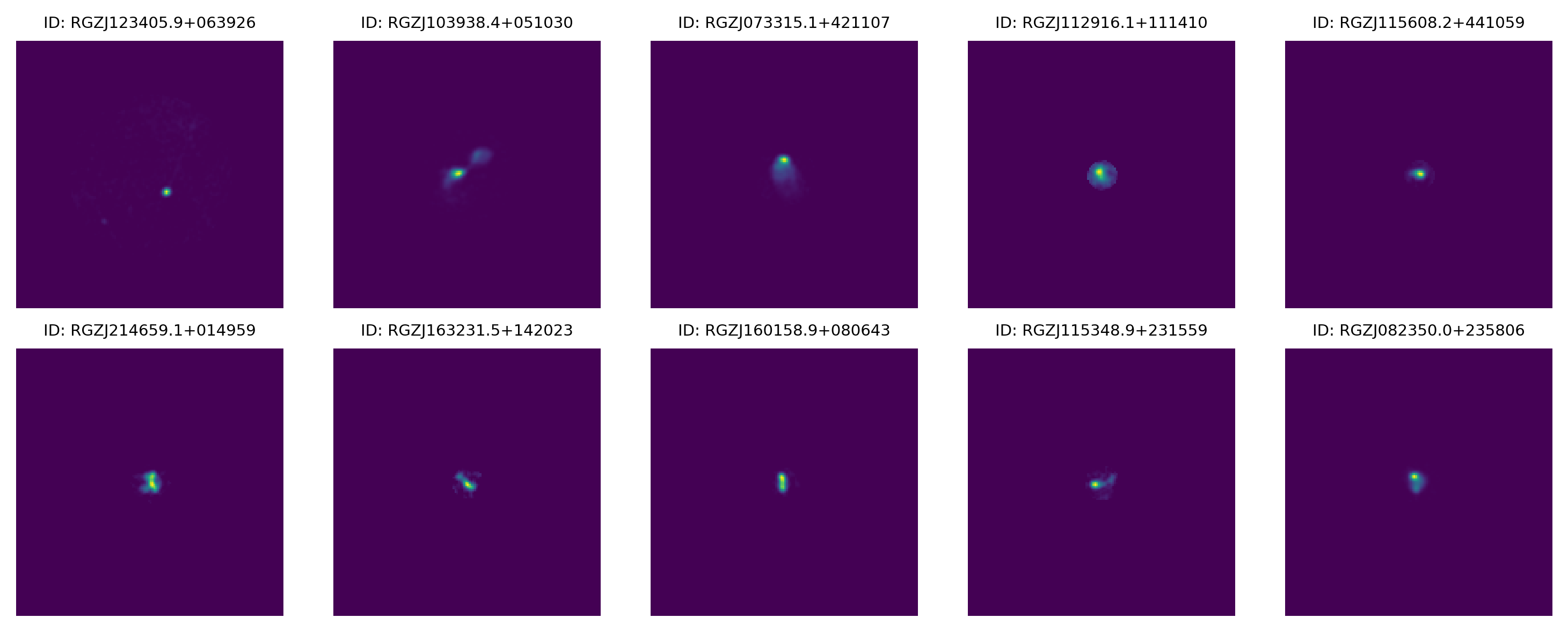}
    \caption{Example radio galaxy cutouts from the RGZ dataset.}
    \label{fig:rgz_cutouts}
\end{figure*}

Early results demonstrated that citizen scientists could achieve accuracy comparable to expert astronomers when consensus levels exceeded 75\%, particularly for simpler morphologies. The majority of identified hosts were found to lie in mid-infrared colour regions typical of elliptical galaxies, quasars, and luminous infrared galaxies (LIRGs), consistent with expectations for radio-loud AGN populations. Furthermore, RGZ enabled the discovery of unusual systems, such as spiral hosts with extended radio lobes and hybrid morphology radio sources (HyMoRS), which are of particular interest for studies of galaxy evolution and AGN feedback \citep{Banfield_2015}.

The scale and diversity of RGZ make it a valuable complement to smaller but curated datasets. While the absence of definitive morphological class labels poses challenges for direct supervised training, the dataset is well-suited for self-supervised representation learning approaches. Its large scale and wide range of source structures allow the development of methods that can better handle the ambiguities and complexities found in real radio galaxy populations.

\section{Model Implementation \& Training}

The VAE implemented in this work adopts a convolutional 
encoder--decoder design tailored to $150 \times 150$ pixel radio galaxy images. 
The encoder consists of stacked convolutional layers with batch normalization 
and ReLU activations, progressively reducing the spatial resolution while 
increasing the number of feature channels. The final feature map is flattened 
and passed through fully connected layers to produce the mean $\mu$ and 
log-variance $\log \sigma^2$ of the approximate posterior distribution. 
Sampling from this distribution is made differentiable using the 
reparameterization trick \citep{kingma2022autoencodingvariationalbayes}.

The decoder reverses this process: latent samples are mapped through fully 
connected layers and reshaped into feature maps, which are then upsampled 
using transposed convolutions. Batch normalization and ReLU activations are 
applied after each deconvolution, and the final output layer uses a sigmoid 
activation to constrain pixel intensities to the $[0,1]$ range.

For the experiments in this thesis, the latent dimensionality is set to 
$z = 8$, providing a compact representation that balances reconstruction 
fidelity with interpretability. A complete layer-by-layer breakdown of the 
architecture is given in Table~\ref{tab:cnnvae_arch}.

This architecture strikes a balance between expressive power and computational efficiency. With a relatively small latent dimension $(z=8)$, the model learns compact representations of radio galaxy morphology that can be used both for reconstruction and as a source of controlled synthetic augmentations.

\begin{table*}
\centering
\caption{CNN-VAE architecture used for experiments ($z=8$). Each layer is listed sequentially.}
\renewcommand{\arraystretch}{1.2}
\setlength{\tabcolsep}{12pt}
\begin{tabular}{l l l l}
\hline
\textbf{Layer} & \textbf{Operation} & \textbf{Parameters} & \textbf{Output Size} \\
\hline
\multicolumn{4}{c}{\textbf{Encoder}} \\
\hline
1 & Convolution & $1 \to 32$, k=4, s=2, p=1 & $32 \times 75 \times 75$ \\
2 & BatchNorm + ReLU & -- & $32 \times 75 \times 75$ \\
3 & Convolution & $32 \to 32$, k=4, s=2, p=1 & $32 \times 38 \times 38$ \\
4 & BatchNorm + ReLU & -- & $32 \times 38 \times 38$ \\
5 & Convolution & $32 \to 64$, k=4, s=2, p=1 & $64 \times 19 \times 19$ \\
6 & BatchNorm + ReLU & -- & $64 \times 19 \times 19$ \\
7 & Convolution & $64 \to 64$, k=4, s=2, p=1 & $64 \times 9 \times 9$ \\
8 & BatchNorm + ReLU & -- & $64 \times 9 \times 9$ \\
9 & Fully Connected & $5184 \to 128$ & 128 \\
10 & Fully Connected (mean) & $128 \to 8$ & $\mu \in \mathbb{R}^8$ \\
11 & Fully Connected (logvar) & $128 \to 8$ & $\log\sigma^2 \in \mathbb{R}^8$ \\
\hline
\multicolumn{4}{c}{\textbf{Decoder}} \\
\hline
1 & Fully Connected & $8 \to 128$ & 128 \\
2 & Fully Connected & $128 \to 5184$ & $64 \times 9 \times 9$ \\
3 & ConvTranspose2D & $64 \to 64$, k=4, s=2, p=1 & $64 \times 19 \times 19$ \\
4 & BatchNorm + ReLU & -- & $64 \times 19 \times 19$ \\
5 & ConvTranspose2D & $64 \to 32$, k=5, s=2, p=1 & $32 \times 38 \times 38$ \\
6 & BatchNorm + ReLU & -- & $32 \times 38 \times 38$ \\
7 & ConvTranspose2D & $32 \to 32$, k=5, s=2, p=1 & $32 \times 75 \times 75$ \\
8 & BatchNorm + ReLU & -- & $32 \times 75 \times 75$ \\
9 & ConvTranspose2D & $32 \to 1$, k=4, s=2, p=1 & $1 \times 150 \times 150$ \\
10 & Sigmoid & -- & $1 \times 150 \times 150$ \\
\hline
\end{tabular}
\label{tab:cnnvae_arch}
\end{table*}

\subsection{Training Setup}
The model is trained to maximize the ELBO with a unit Gaussian prior $p(z) = \mathcal{N}(0,I)$. The reconstruction term is implemented as binary cross-entropy (BCE) under a Bernoulli likelihood, while the KL term is weighted by a factor $\beta$. To stabilize training, KL annealing is applied where
\begin{equation}
    \beta_t = \min\!\left(1, \frac{t+1}{10}\right)\,\beta_{\max},
\end{equation}
such that $\beta$ increases linearly over the first $t=10$ epochs before being fixed at $\beta_{\max}$.

Training is fully unsupervised, with inputs normalized to $[0,1]$. Optimization uses Adam with learning rate $10^{-4}$ and runs for 150 epochs. MiraBest and RGZ images are provided as $150 \times 150$ cutouts.

\section{BYOL Framework}
This work employs the BYOL framework \citep{grill2020byol} as the primary self-supervised learning method for radio galaxy images. BYOL was selected due to its proven ability to learn high-quality feature representations without requiring negative samples, making it particularly well suited for domains with limited labels but abundant unlabelled data, such as radio astronomy. The implementation used here follows the adaptation of BYOL to astronomical datasets presented in \cite{slijepcevic2023radiogalaxyzoobuilding}, ensuring consistency with recent work on large-scale self-supervised modeling of radio galaxies. The complete implementation of our pipeline, including the $\beta$-VAE and augmentation modules, is made publicly available to facilitate reproducibility.\footnote{The code repository is available at: \url{https://github.com/joe-johnny/b_VAE-for-RadioGalaxy-SSL}}

The following subsections outline the specific components of this setup: the network architecture used for encoding and projection, the augmentation pipeline applied to generate diverse training views, and the evaluation protocol adopted to assess representation quality.

\subsection{Network Architecture}
The network architecture adopted in this thesis follows the design implemented by \cite{slijepcevic2023radiogalaxyzoobuilding} for applying BYOL to radio galaxy data. Both the online and target networks use a ResNet-18 backbone \citep{he2015deepresiduallearningimage}, chosen after a model-depth comparison showed little improvement from deeper variants such as ResNet-34. This choice provides a good balance between representational capacity and computational efficiency for single-channel, structurally sparse radio images. The ResNet encoder processes input cutouts of $128 \times 128$ pixels, extracting spatial features that form the basis for the downstream projection and prediction modules used in the BYOL framework.

On top of this encoder, BYOL introduces lightweight MLPs. Both branches include a projection head, which maps encoder outputs into a latent space suited for the self-supervised objective. In the online branch, an additional predictor head is applied, providing the asymmetry that enables BYOL to avoid representational collapse. The target branch mirrors the encoder and projector of the online network but omits the predictor. Its parameters are not directly updated by gradients; instead, they are maintained as an EMA of the online parameters, which stabilizes training and ensures the target representations evolve smoothly.

\begin{table}[h!]
\centering
\caption{BYOL architecture, following \citet{slijepcevic2023radiogalaxyzoobuilding}}
\renewcommand{\arraystretch}{1.2} % row spacing
\setlength{\tabcolsep}{6pt} % column spacing
\begin{tabular}{l c c c}
\hline
\textbf{Module} & \textbf{Online} & \textbf{Target} & \textbf{Output Dim.} \\
\hline
Encoder         & ResNet-18 & ResNet-18 & 512 \\
Projection Head & 2-layer MLP & 2-layer MLP & 256 \\
Predictor Head  & 2-layer MLP & -- & 256 \\
Update Rule     & Backprop & EMA ($\tau=0.99$) & -- \\
\hline
\end{tabular}
\label{tab:byol_architecture}
\end{table}

The overall structure of the BYOL framework is summarized in Table~\ref{tab:byol_architecture}, which highlights the symmetry between the online and target branches, as well as the role of the predictor head in maintaining asymmetry. The associated training hyperparameters, shown in Table~\ref{tab:byol_hyperparameters}.

\begin{table}[h!]
\centering
\caption{Training hyperparameters for BYOL pretraining, following \citet{slijepcevic2023radiogalaxyzoobuilding}.}
\renewcommand{\arraystretch}{1.3} % row spacing
\setlength{\tabcolsep}{12pt} % column spacing
\begin{tabular}{l c}
\hline
\textbf{Hyperparameter} & \textbf{Value / Setting} \\
\hline
Optimizer                & SGD with momentum 0.9 \\
Weight Decay             & $1.5 \times 10^{-6}$ \\
Base Learning Rate       & 0.2 (scaled with batch size) \\
Batch Size               & 512 \\
Learning Rate Schedule   & 10-epoch warm-up + cosine decay \\
Epochs                   & 800 \\
Input Resolution         & $128 \times 128$ pixels (single-channel) \\
\hline
\end{tabular}
\label{tab:byol_hyperparameters}
\end{table}

During pretraining, all components work together to align the online predictions with the target projections of differently augmented views of the same image. For downstream tasks such as classification, only the encoder is kept, while the projection and prediction heads are discarded. This ensures the final model is compact and directly comparable to supervised baselines.

\subsection{Augmentation Pipeline}
The augmentation pipeline is central to the effectiveness of BYOL, as it generates multiple distinct “views” of the same input source to enforce invariance in the learned representations. In this work, the design follows the pipeline proposed by \citet{slijepcevic2023radiogalaxyzoobuilding}, with the addition of $\beta$-VAE reconstructions introduced as an augmentation stage prior to conventional transformations, as summarised in Table \ref{tab:byol_augmentations}. Unlike entirely synthetic generations, these $\beta$-VAE reconstructions retain the overall morphology of the original input while introducing stochastic morphological variations that reflect the disentangled structure of the learned latent space. Concretely, each input image is passed through the $\beta$-VAE encoder to obtain the posterior mean, $\mu$, and log-variance, $\log{\sigma}^2$. A variate is then sampled from the latent space using the reparameterization trick across all 8 latent dimensions simultaneously, such that $\mathbf{z} = \mu + \sigma \odot \epsilon$, where $\epsilon \sim \mathcal{N} (0, 0.5\,I)$, and decoded to produce the augmented view. The sampling variance of 0.5 was chosen empirically as it introduces meaningful morphological diversity while preserving the overall source structure. This provides a complementary and semantically meaningful view of the same radio galaxy, encouraging the network to learn representations that are invariant not only to geometric transformations but also to morphological diversity.

Following this reconstruction step, each image undergoes a sequence of standard augmentations. Inputs are randomly cropped and resized to 128×128 pixels (down from the original 150×150 cutouts), introducing positional variability. Horizontal and vertical flips are applied to exploit the approximate rotational invariance of radio galaxy morphologies. To further enhance robustness, Gaussian blurring is used to mimic resolution degradation, while Color Jitter introduces changes in contrast and saturation.

\begin{table}[h!]
\centering
\caption{Augmentations used in BYOL training. Standard transformations follow \citet{slijepcevic2023radiogalaxyzoobuilding}, with $\beta$-VAE reconstructions added as an additional view.}
\renewcommand{\arraystretch}{1.2}
\setlength{\tabcolsep}{12pt}
\begin{tabular}{l l}
\hline
\textbf{Augmentation} & \textbf{Value} \\
\hline
$\beta$-VAE reconstruction   & Re-encode + decode of input \\
Rotation             & 0 -- 360° (uniform distribution) \\
Center crop          & 70 pixels \\
Random resized crop  & 80 -- 100\% of original size \\
Horizontal flip      & $p = 0.5$ \\
Vertical flip        & $p = 0.5$ \\
Color jitter         & $s = 0.5$, $p = 0.8$ \\
Gaussian blur        & $p = 0.1$ \\
\hline
\end{tabular}
\label{tab:byol_augmentations}
\end{table}

By integrating $\beta$-VAE reconstructions into the augmentation pipeline alongside traditional geometric and photometric transforms, this approach introduces both structural and observational variability into the learning process. Such a hybrid strategy is particularly well-suited to radio astronomy data, where models must be resilient to both imaging artifacts and the complex, continuous nature of source morphology.

\subsection{Finetuning and Evaluation}
\label{ch:eval}
To assess the quality of representations learned through BYOL pretraining, we adopt two complementary evaluation protocols: linear evaluation and fine-tuning.
In the linear evaluation setup, commonly used as a standard benchmark in self-supervised learning \citep{grill2020byol}, the encoder is frozen after pretraining, and a single fully connected classifier is trained on top using the MiraBest dataset \citep{porter2023mirabestdatasetmorphologicallyclassified}. This approach directly measures the quality of the learned representations, as no further updates are made to the encoder weights. The classification performance thus reflects how well the features generalize to a supervised task without additional adaptation.

In contrast, the fine-tuning setup unfreezes the encoder, allowing all layers to be jointly trained with the classifier on labelled MiraBest data. This approach evaluates not only the quality but also the adaptability of the pretrained encoder when exposed to limited supervision. Fine-tuning can often uncover representational nuances that remain latent under linear probing, especially when domain shifts or label-specific features are involved.

Following \citet{slijepcevic2023radiogalaxyzoobuilding}, both protocols are evaluated using binary classification of Fanaroff-Riley types (FR I vs. FR II), the canonical task in radio galaxy morphology. Accuracy is reported as the primary metric, supplemented by class-wise performance where relevant to address potential imbalances in the dataset.

Additionally, ablation studies are conducted under varied augmentation configurations, standard-only, $\beta$-VAE-only, and hybrid, allowing for a more detailed understanding of how augmentation design and generative views impact both representation quality and transferability.

\section{Discussions}
\label{sec:discussion}
\subsection{$\beta$-VAE Representation Learning}

\subsubsection{Selection of Regularization Strength ($\beta$)}
Quantitative evaluation of disentanglement remains an open challenge in unsupervised learning. Standard metrics such as the $\beta$-VAE metric \citep{Higgins2016betaVAELB} or FactorVAE metric \citep{kim2019disentanglingfactorising} require ground-truth labels for the generative factors of variation, which are unavailable for real observational data. Furthermore, recent studies suggest that these metrics can be unstable and do not always correlate with downstream task performance \citep{locatello2019challengingcommonassumptionsunsupervised}. Consequently, we adopted a heuristic approach combining visual inspection with a "capacity diagnostic" analysis to determine the optimal $\beta$.

We monitored the trade-off between the KL divergence (regularization load) and the mean log-variance of the latent posterior (a proxy for information capacity) across a range of $\beta$ values. The intersection of these metrics typically indicates a balance point where the model effectively compresses data without suffering from posterior collapse.

For the RGZ dataset, which served as the generator for our downstream BYOL pipeline, this diagnostic (Fig. \ref{fig:rgz_diag}) revealed a balance point at approximately $\beta \approx 2.3$. This quantitative finding aligned with our visual inspection, where $\beta = 2.3$ produced the most coherent morphologies, effectively smoothing out the noise inherent in the dataset while retaining structural diversity. Given this convergence of quantitative and qualitative evidence, this model was selected to generate the augmentations for the self-supervised learning experiments.

\begin{figure}
    \includegraphics[width=0.5\textwidth]{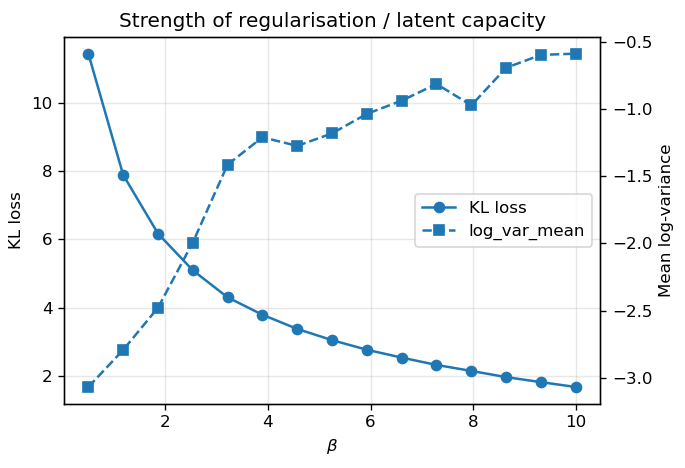}
    \caption{Diagnostic plot of Regularization Strength ($\beta$) vs. Latent Capacity for the RGZ dataset. The balance point where the curves intersect aligns with the empirically chosen $\beta=2.3$.}
    \label{fig:rgz_diag}
\end{figure}

\subsubsection{Reconstruction and Generation Behavior}
With the regularization parameters fixed at 
%$\beta=1.22$ (MiraBest) and 
$\beta=2.3$, we evaluated the qualitative fidelity of the model outputs. The visual results confirm that this setting achieved the necessary balance between structural definition and noise suppression. In our analysis, models trained with significantly lower $\beta$ values produced sharper images but frequently reconstructed background artifacts, while those with higher $\beta$ values yielded over-smoothed morphologies that lost astrophysical details such as hotspots.

Reconstructions with $\beta$ = 2.3 (Fig.~\ref{fig:rgz_23_rec}) maintained the essential two-component FR II morphology while allowing controlled variability. The model reliably reproduced symmetric double lobes and their orientation, with differences emerging gradually as variance increased. Generated sources (Fig.~\ref{fig:rgz_23_gen}) exhibited a continuum of compact to extended morphologies, including mild brightness asymmetries and diffuse emission consistent with physical variability across the population.

Overall, the $\beta$-VAE demonstrates that moderate $\beta$ values yield latent spaces capable of producing morphologically stable yet semantically flexible reconstructions and generations. Variance governs the degree of diversity within these spaces, slow variance ensures fidelity to the input, while higher variance explores plausible structural variability, confirming that the $\beta$-VAE framework effectively captures the generative continuum of radio galaxy morphologies.

\begin{figure*}
\centerline{\includegraphics[width=0.85\textwidth]{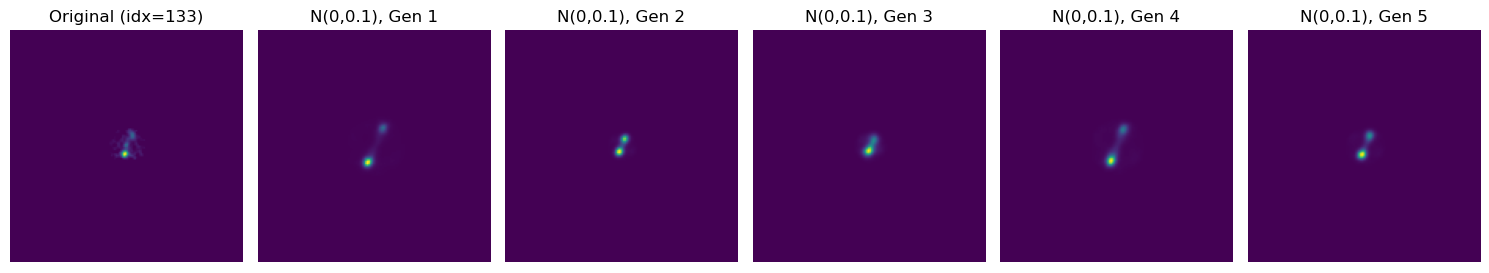}}    
\centerline{(a)}
\centerline{\includegraphics[width=0.85\textwidth]{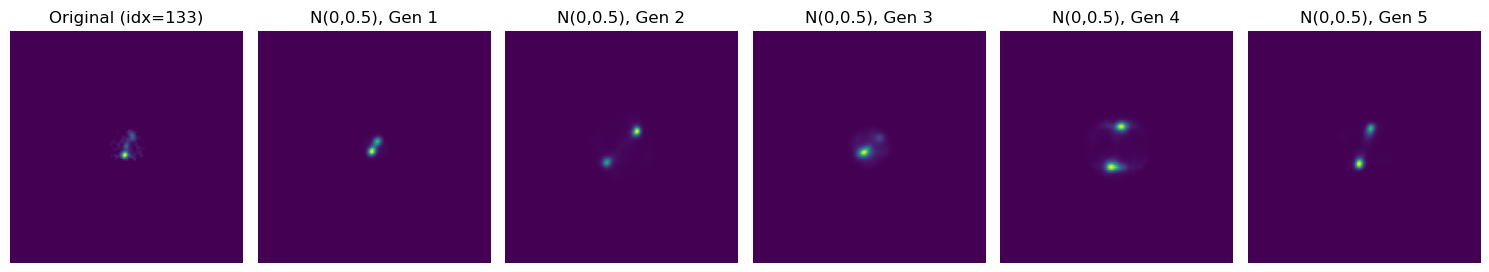}} 
\centerline{(b)}
\centerline{\includegraphics[width=0.85\textwidth]{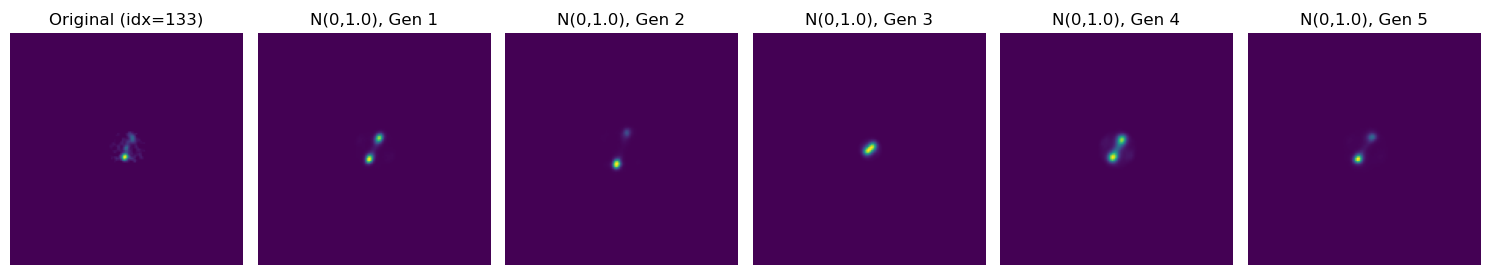}}
\centerline{(c)}
 \caption{Reconstructions of a RGZ radio source with a $\beta$-VAE \textbf{($\beta = 2.3$)}. The original is shown at left in each row, followed by multiple stochastic reconstructions obtained by adding Gaussian noise to the latent code at increasing levels: (A)$\sigma^2 = 0.1$, (B)$\sigma^2 = 0.5$, (C) $\sigma^2 = 1.0.$ 
 % (A) $\mathcal{N}(0,0.1)$, (B) $\mathcal{N}(0,0.5)$, (C) $\mathcal{N}(0,1.0)$.
 }
     \label{fig:rgz_23_rec}
\end{figure*}

\begin{figure*}
\centerline{\includegraphics[width=0.85\textwidth]{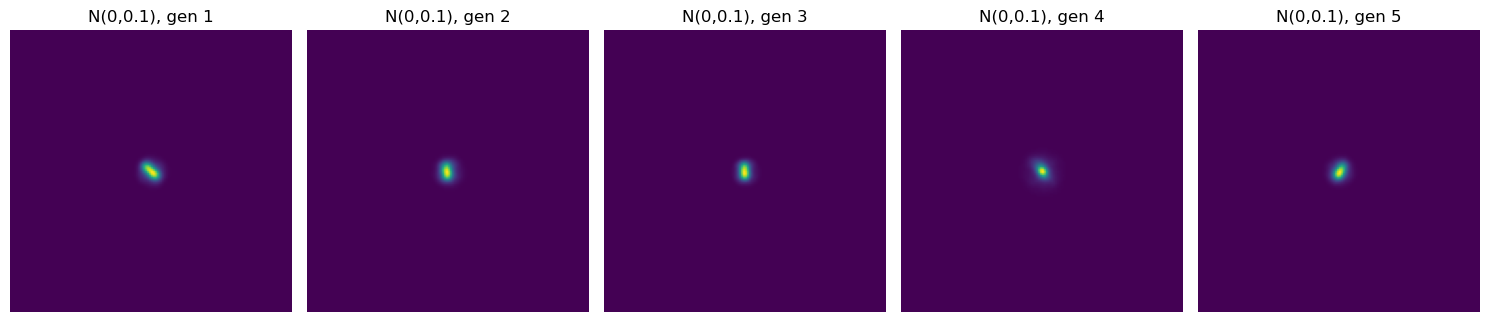}}    
\centerline{(a)}
\centerline{\includegraphics[width=0.85\textwidth]{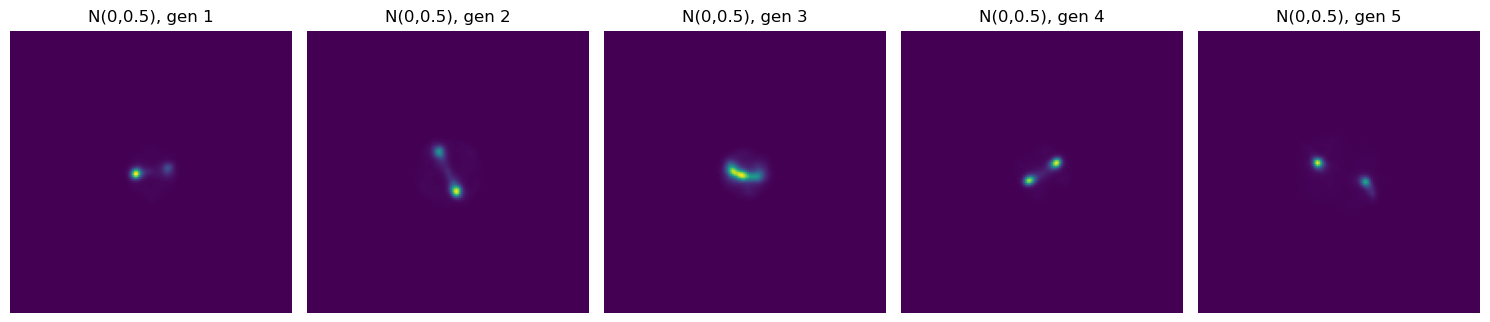}} 
\centerline{(b)}
\centerline{\includegraphics[width=0.85\textwidth]{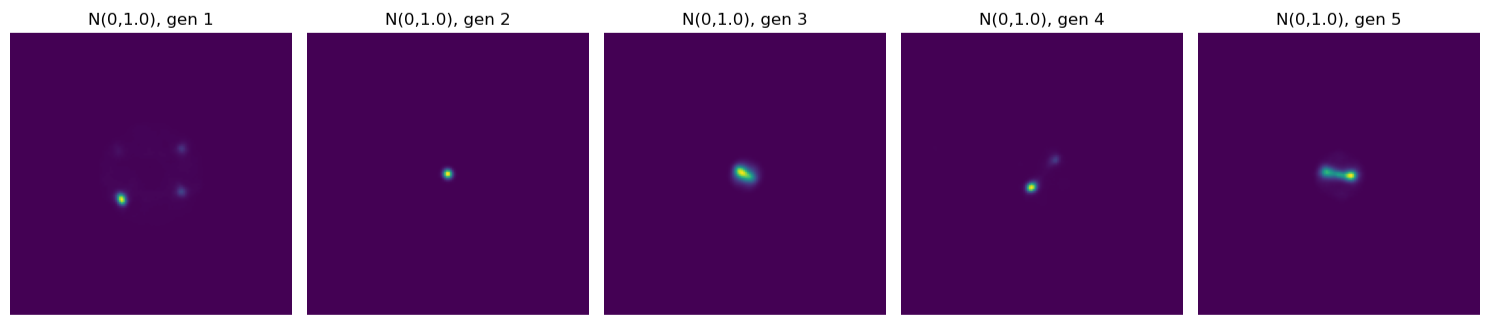}}
\centerline{(c)}
 \caption{Synthetic samples of radio sources from the latent space of a $\beta$-VAE trained on the Radio Galaxy Zoo dataset with $\beta = 2.3$. Each row shows generations obtained by adding Gaussian noise to the latent code at increasing levels: (A)$\sigma^2 = 0.1$, (B)$\sigma^2 = 0.5$, (C) $\sigma^2 = 1.0.$ 
 % (A) $\mathcal{N}(0,0.1)$, (B) $\mathcal{N}(0,0.5)$, and (C) $\mathcal{N}(0,1.0)$.
 }
    \label{fig:rgz_23_gen}
\end{figure*}

\subsubsection{Latent Structure and Disentanglement}
Latent traversal and projection analyses were used to provide insight into the internal structure of the $\beta$-VAE representations. In Figure \ref{fig:LT_rgz_23}, each row corresponds to one of the 8 latent dimensions, $z_i$, and the columns show the decoded output as that dimension is varied from -3.0 to +3.0 standard deviations while all other dimensions are held fixed at zero. It was observed that the chosen $\beta$ value encouraged meaningful disentanglement of several morphological factors.
These include variations in core–lobe separation, lobe luminosity asymmetry, elongation, halo or jet-like diffusion, and orientation. Each factor changes gradually and independently, producing smooth transitions that align with plausible physical variations seen in radio galaxies. Some latent dimensions, however, remain less interpretable, showing irregular or diffuse patterns that may encode residual noise or blended features rather than distinct morphological factors (see Figure~\ref{fig:LT_rgz_23}). The latent space clearly encodes lobe brightness asymmetry, orientation and axis rotation, component multiplicity, and lobe separation. These disentangled behaviours indicate that the model organizes morphology along continuous axes rather than discrete class boundaries, providing a smooth manifold of structural transformations.

The UMAP projection of the RGZ latent space, shown in Figure~\ref{fig:rgz_umap}, reinforces this observation. Rather than forming distinct clusters, the projection reveals a continuous gradient in which FR I and FR II sources occupy overlapping but ordered regions. This is consistent with the known astrophysical continuum between these classes \citep{headtail, gopalkrishna2000extragalacticradiosourceshybrid}, and reflects the fact that the $\beta$-VAE is optimized for faithful reconstruction rather than class separability. The latent space therefore organizes radio galaxy morphology along continuous physical axes, providing the well-structured generative manifold required for producing meaningful stochastic augmentations in the BYOL pretraining pipeline.

\begin{figure*}
    \centerline{\includegraphics[width=0.95\textwidth]{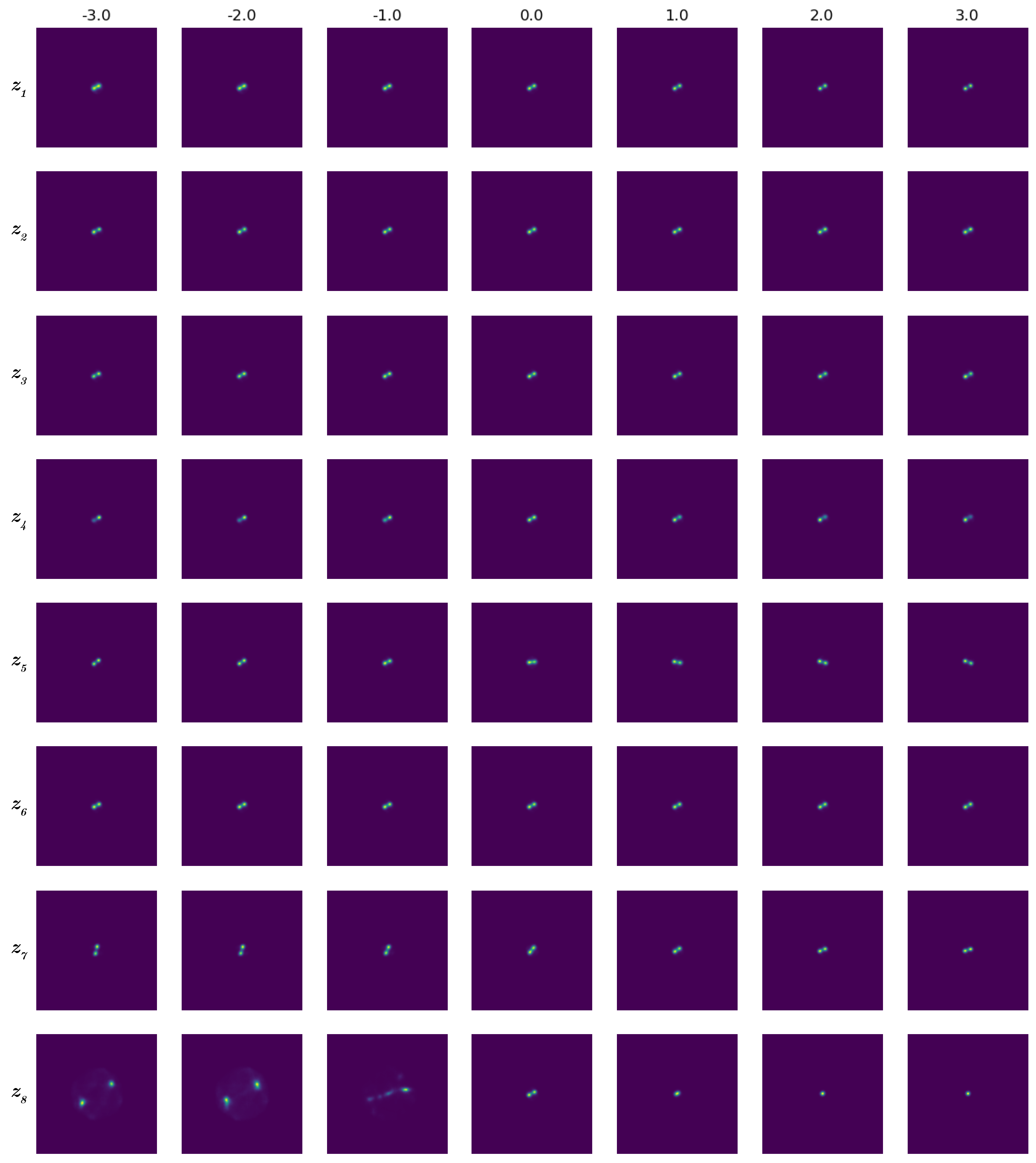}}
    \caption{Latent traversal map showing all 8 dimensions with disentangled factors in RGZ Data using $\beta$-VAE with \textbf{$\beta = 2.3$}}
    \label{fig:LT_rgz_23}
\end{figure*}

\begin{figure}
    \includegraphics[width=0.4\textwidth]{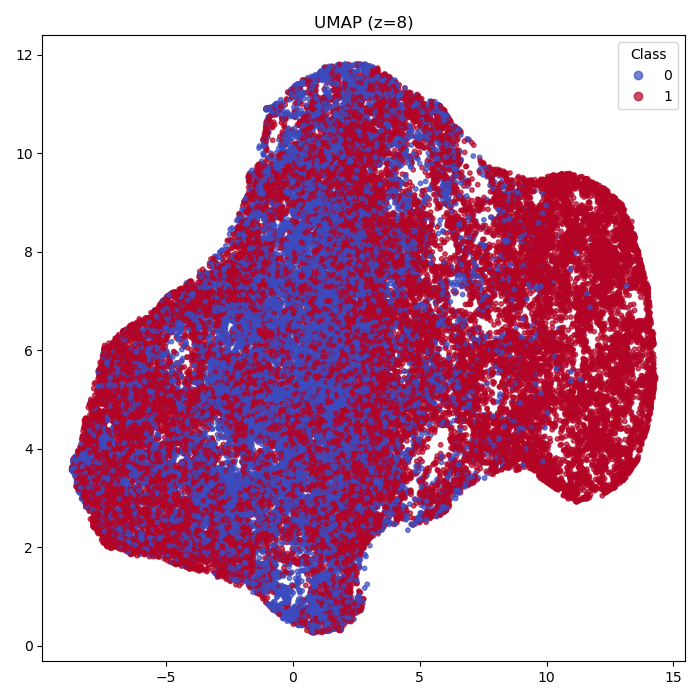}
    \caption{UMAP projection of the $z=8$ latent space learned by the standard VAE on the RGZ dataset. t. Points are coloured by FR class label (blue = FR I, red = FR II) derived from \citep{Buatthaisong_2025} for a subset of RGZ sources, used only for visualization.}
    \label{fig:rgz_umap}
\end{figure}

\subsection{BYOL Ablation Study}
\label{sec:Byol_results}
To quantify the effect of individual transformations in the BYOL pipeline, two ablation regimes were tested:
\begin{enumerate}
    \item a no $\beta$-VAE setup, using only standard augmentations (rotation, crop, flip, color jitter, and Gaussian blur) and

    \item a +$\beta$-VAE setup, where $\beta$-VAE reconstructions, generated from a $\beta$-VAE trained on the unlabelled RGZ dataset at $\beta$ = 2.3, were incorporated as additional generative views.
\end{enumerate}

\begin{figure}
    \includegraphics[width=0.47\textwidth]{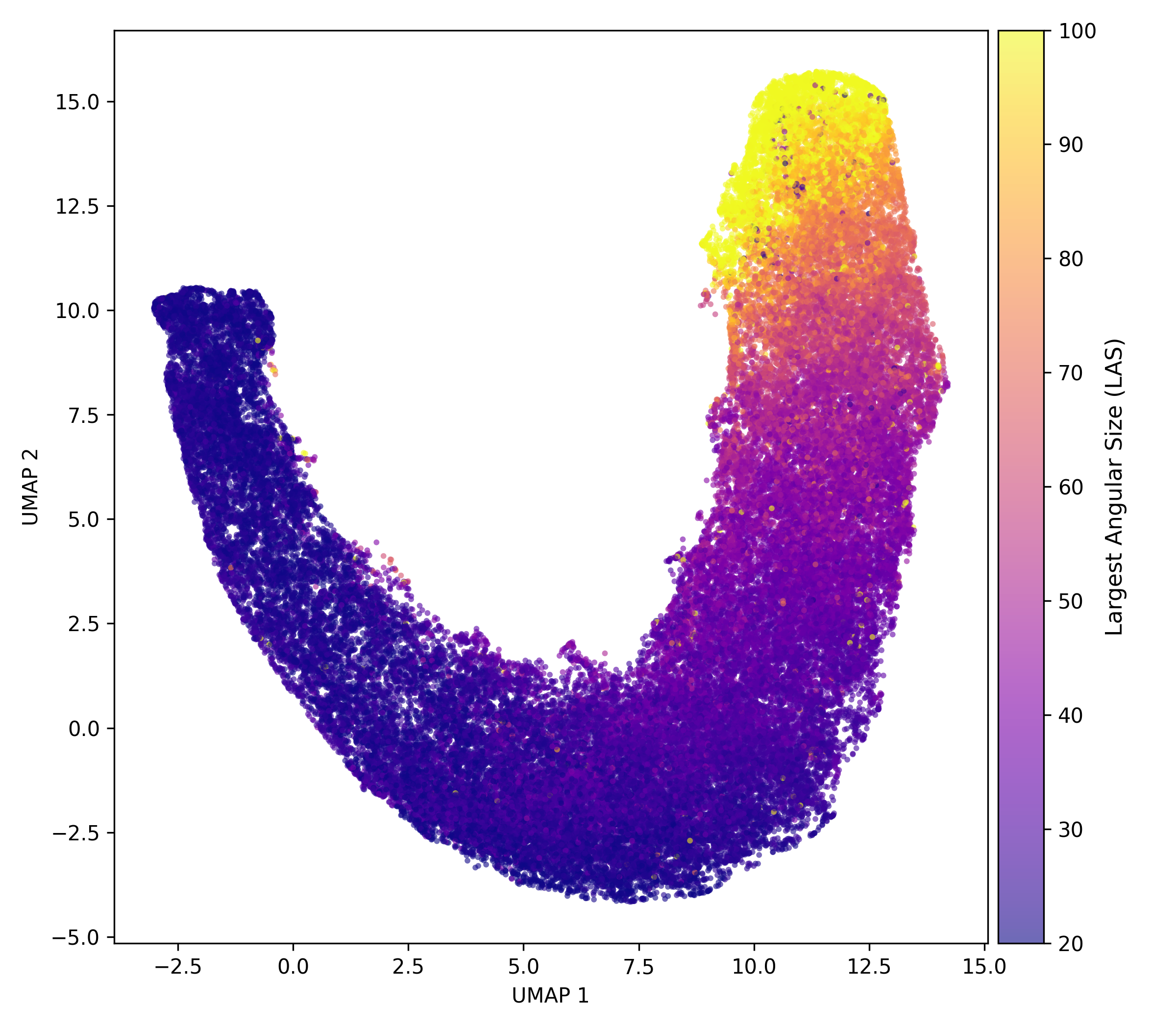}
    \caption{UMAP projection of the RGZ dataset passed through the pretrained BYOL encoder trained with all augmentations (All augmentations + $\beta$-VAE). Points are coloured by the largest angular size (LAS) of each source in arcseconds. The continuous gradient from compact (purple) to extended (yellow/orange) sources demonstrates that the encoder has learned morphologically meaningful representations of radio galaxy structure.}
    \label{fig:byol_umap}
\end{figure}

\begin{figure*}
    \centerline{\includegraphics[width=0.95\textwidth]{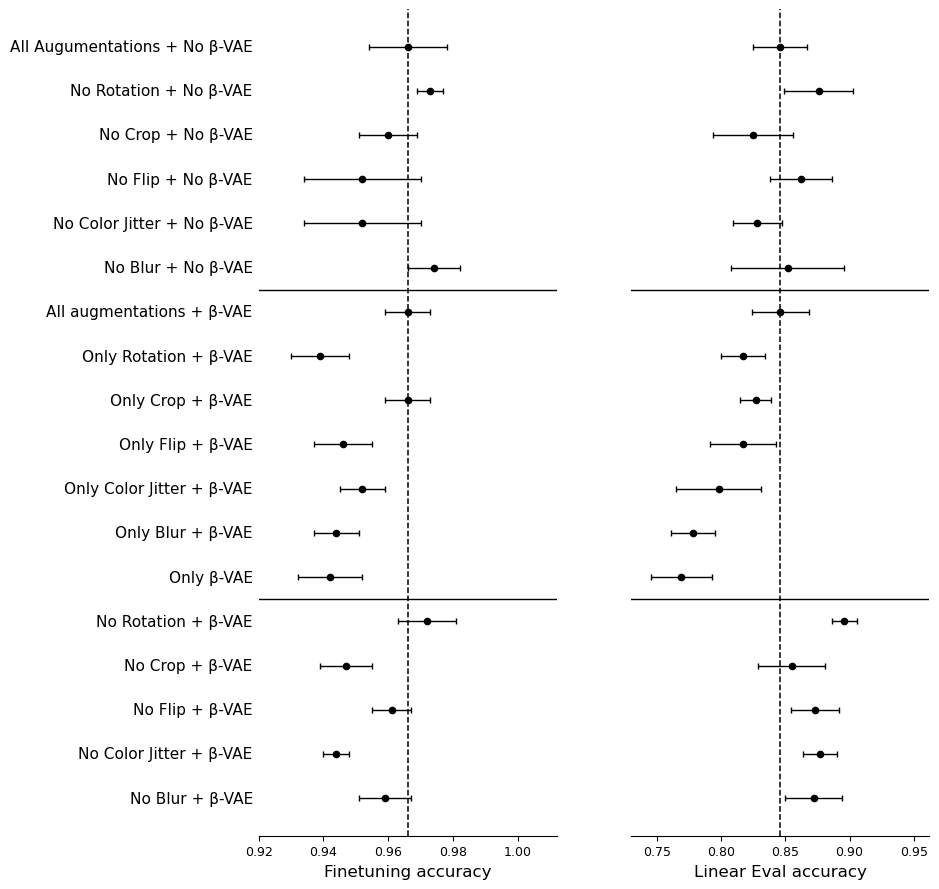}}
    \caption{Performance of BYOL with and without $\beta$-VAE under different augmentation ablations on the MiraBest Confident dataset is shown. The x axes for the finetuning and linear evaluation are scaled differently to show the relative difference between each value. Numerical results shown in Table~\ref{tab:byol_num}.}
    \label{fig:byol_results_table}
\end{figure*}

\begin{table*}
\centering
\renewcommand{\arraystretch}{1.3} % row spacing
\setlength{\tabcolsep}{8pt} % column spacing
\caption{Performance with and without $\beta$-VAE under different augmentation ablations on the MiraBest Confident dataset, reported as mean $\pm$ standard deviation across 10 independent runs with different seeds. $\checkmark$ indicates augmentation applied, $\times$ indicates removed. See Figure \ref{fig:byol_results_table} for a visual summary.}
\label{tab:byol_num}
\begin{tabular}{c c c c c c c c}
\hline
Rotation & Crop & Flip & Color Jitter & Blur & $\beta$-VAE & Finetuning & Linear Eval \\ \hline
\checkmark & \checkmark & \checkmark & \checkmark & \checkmark & $\times$ &  $0.966 \pm 0.012$ & $0.846 \pm 0.021$ \\
$\times$ & \checkmark & \checkmark & \checkmark & \checkmark & $\times$ &  $0.973 \pm 0.004$ & $0.876 \pm 0.027$ \\
\checkmark & $\times$ & \checkmark & \checkmark & \checkmark & $\times$ & $0.960 \pm 0.009$ & $0.825 \pm 0.031$ \\
\checkmark & \checkmark & $\times$ & \checkmark & \checkmark & $\times$ & $0.952 \pm 0.018$ & $0.862 \pm 0.024$ \\
\checkmark & \checkmark & \checkmark & $\times$ & \checkmark & $\times$ & $0.952 \pm 0.018$ & $0.828 \pm 0.019$ \\
\checkmark & \checkmark & \checkmark & \checkmark & $\times$ & $\times$ & $0.974 \pm 0.008$ & $0.852 \pm 0.044$ \\ 
\hline
\checkmark & \checkmark & \checkmark & \checkmark & \checkmark & \checkmark & $0.966 \pm 0.007$ & $0.846 \pm 0.022$ \\
\checkmark     & $\times$ & $\times$     & $\times$       & $\times$ & \checkmark & $0.939 \pm 0.009$ & $0.817 \pm 0.017$ \\
$\times$  &  \checkmark  & $\times$     & $\times$   & $\times$ & \checkmark & $0.966 \pm 0.007$ & $0.827 \pm 0.012$ \\
$\times$ & $\times$ & \checkmark     & $\times$       & $\times$ & \checkmark & $0.946 \pm 0.009$ & $0.817 \pm 0.026$ \\
$\times$ & $\times$ & $\times$ & \checkmark       & $\times$ & \checkmark & $0.952 \pm 0.007$ & $0.798 \pm 0.033$ \\
$\times$ & $\times$ & $\times$ & $\times$  & \checkmark & \checkmark & $0.944 \pm 0.007$ & $0.778 \pm 0.017$ \\
$\times$ & $\times$ & $\times$ & $\times$   & $\times$ & \checkmark & $0.942 \pm 0.010$ & $0.769 \pm 0.024$ \\ \hline
$\times$     & \checkmark & \checkmark & \checkmark   & \checkmark & \checkmark & $0.972 \pm 0.009$ & $0.896 \pm 0.010$ \\
\checkmark & $\times$     & \checkmark & \checkmark   & \checkmark & \checkmark &  
$0.947 \pm 0.008$ & $0.855 \pm 0.026$\\
\checkmark & \checkmark & $\times$     & \checkmark   & \checkmark & \checkmark & $0.961 \pm 0.006$ & $0.873 \pm 0.019$ \\
\checkmark & \checkmark & \checkmark & $\times$       & \checkmark & \checkmark & $0.944 \pm 0.004$ & $0.877 \pm 0.013$ \\
\checkmark & \checkmark & \checkmark & \checkmark   & $\times$ & \checkmark & $0.959 \pm 0.008$ & $0.872 \pm 0.022$ \\ \hline
\end{tabular}
\end{table*}

In the no $\beta$-VAE setup, results show that random crop and color jitter are the most critical augmentations for learning robust representations. Removing crop caused the largest performance drop, confirming its role in enforcing positional and scale invariance by focusing attention on the galaxy region rather than background noise. Excluding color jitter also reduced accuracy, indicating its importance in teaching the model stability to natural brightness and flux variations. Conversely, removing rotation slightly improved results, suggesting that strict rotational invariance may obscure meaningful directional cues in FR morphology. Flipping and blurring had smaller impacts; the latter even slightly hindered performance by removing key spatial details such as edges and hotspots.

In the $+\beta$-VAE setup, a direct comparison of the fully augmented baselines reveals that adding $\beta$-VAE views to the standard pipeline yields performance comparable to the 'No $\beta$-VAE' setup. This suggests that the inclusion of generative views does not immediately overcome the limitations imposed by formal geometric transformations, such as rotation, when all are applied simultaneously. However, the contribution of the $\beta$-VAE becomes evident in the ablations. When used in isolation (“Only $\beta$-VAE”), performance declined, indicating that synthetic views alone cannot replace standard transformations. Among the ablations, removing rotation again yielded the strongest improvement (linear = 0.896 $\pm$ 0.010), implying that orientation diversity is already captured within the $+\beta$-VAE latent space. Similarly, removing color jitter slightly improved linear performance, suggesting that the flux variation encoded in $+\beta$-VAE reconstructions reduces the need for photometric noise. Crop remained essential, as its removal led to reduced linear accuracy, confirming its continued importance for localization and scale invariance.

Figure~\ref{fig:byol_umap} shows a UMAP projection of the RGZ dataset passed through the pretrained encoder trained with all augmentations, confirming that the learned representations encode morphologically meaningful structure, with sources organised continuously by angular extent. A visual summary of the ablation results is shown in Figure~\ref{fig:byol_results_table}, with numerical results in Table~\ref{tab:byol_num}. In both the figure and table, each point represents the mean accuracy across 10 independent training runs with seeds 0–9, and the error bars show the corresponding standard deviation.

These ablation results indicate that $\beta$-VAE reconstructions serve as effective, morphology preserving augmentations that strengthen BYOL’s self-supervised learning process. When paired with key geometric transformations, especially cropping and flipping, they improve representation quality while reducing reliance on more aggressive photometric or rotational perturbations. The findings demonstrate that generative and contrastive approaches can work synergistically to produce semantically rich and astrophysically meaningful embeddings of radio galaxy morphology.

\section{Conclusions}
In this work, we have presented a novel self-supervised learning framework for radio astronomy that integrates generative views from a $\beta$-VAE into the BYOL pipeline. Motivated by the unprecedented scale of data expected from next-generation surveys like the SKA, we investigated whether disentangled generative factors could serve as semantically meaningful augmentations to improve representation learning without reliance on extensive labelled catalogues.

Our analysis of the $\beta$-VAE component demonstrated that moderate regularization ($\beta=2.3$ for RGZ) effectively disentangles key morphological factors such as lobe separation, asymmetry, and source orientation. Crucially, we found that the latent spaces of these models do not align with discrete Fanaroff-Riley class boundaries but rather encode morphology as a continuous manifold. This supports the growing consensus that radio galaxy morphology is better described by continuous physical parameters than by binary classification schemes \citep{headtail, gopalkrishna2000extragalacticradiosourceshybrid}.

The ablation studies reveal a clear hierarchy in the effectiveness of augmentation strategies. We found that relying only on generative views is detrimental to representation learning. Performance declined notably when the $\beta$-VAE was used in isolation, demonstrating that synthetic variations cannot yet replace the spatial priors provided by standard geometric transformations, particularly cropping. However, the generative model serves as a valuable complement to these traditional augmentation methods. The $\beta$-VAE reconstructions effectively compensated for the removal of rotational and photometric augmentations, suggesting that the model captures intrinsic orientation and flux variability. This indicates that while generative models cannot yet serve as a standalone augmentation strategy, they function as a vital auxiliary mechanism that incorporates the astrophysical structure into the self-supervised pipeline. Ultimately, this work highlights the potential of disentanglement-aware self-supervised learning, paving the way for foundation models that are not only robust to observational artifacts but are also physically informative.

\section*{Acknowledgements}

AMS gratefully acknowledges support from the UK Alan Turing Institute under grant reference EP/V030302/1.

%%%%%%%%%%%%%%%%%%%%%%%%%%%%%%%%%%%%%%%%%%%%%%%%%%
\section*{Data Availability}

The data underlying this article were derived from sources in the public domain: the MiraBest dataset \citep{porter2023mirabestdatasetmorphologicallyclassified} and the Radio Galaxy Zoo dataset \citep{Banfield_2015}. The code underlying this article is available in the GitHub repository at \url{https://github.com/joe-johnny/b_VAE-for-RadioGalaxy-SSL}.

\section*{Conflict of interest}
The authors declare no conflict of interest.

%%%%%%%%%%%%%%%%%%%% REFERENCES %%%%%%%%%%%%%%%%%%

% The best way to enter references is to use BibTeX:

\bibliographystyle{mnras}
\bibliography{vae_byol} 

%%%%%%%%%%%%%%%%%%%%%%%%%%%%%%%%%%%%%%%%%%%%%%%%%%

% Don't change these lines
\bsp	% typesetting comment
\label{lastpage}
\end{document}